\documentclass[12pt]{article}
\usepackage{amsmath}
\usepackage{amsfonts}
\usepackage{amssymb}
\usepackage{srcltx}

\def\oi{b_{\infty }}

\def\gli{gl_{\infty }}

\newtheorem{theorem}{Theorem}
\newtheorem{proposition}[theorem]{Proposition}
\newtheorem{lemma}[theorem]{Lemma}

\newtheorem{remark}[theorem]{Remark}

\begin{document}
\noindent
{\Large {\bf BKP tau-functions as square roots of KP tau-functions }}
\vskip 9 mm
\begin{minipage}[t]{70mm}
{\bf Johan van de Leur}\\
\\
Mathematical Institute,\\
Utrecht University,\\
P.O. Box 80010, 3508 TA Utrecht,\\
The Netherlands\\
e-mail: J.W.vandeLeur@uu.nl
\end{minipage}
\\
\
\\
\begin{abstract}
It is well-known that a BKP tau-function is the square root of a certain KP tau-function, provided one puts the even KP  times equal to zero. In this paper we compute for all polynomial BKP tau-function its corresponding  KP "square". We also give, in the polynomial case, a representation theoretical proof of a recent result by Alexandov, viz. that a KdV tau-function becomes a BKP tau-function when one divides all KdV times by 2.
\end{abstract}
\section{Introduction}
In the 1980's Date, Jimbo, Kashiwara and Miwa, inspired by the pioneering work of Sato \cite{S}, described many soliton hierarchies of KP  and KdV type \cite{DJKM1}, \cite{DJKM2}, \cite{DJKM3}, \cite{JM}. In particular they introduced the BKP hierarchy in \cite{DJKM3}, which is related to the lie algebra $b_\infty$.  They define the level one spin module of this infinite dimensonal orthogonal Lie algebra by action of certain fermionic creation and annihilation operators on a vacuum vector $|0\rangle $. The BKP hierarchy describes the $B_\infty$-group orbit of this highest weight vector  $|0\rangle $ in this spin module. Elements in this orbit are  the BKP tau-functions, which, in the polynomial case,  can be  describe as certain Pfaffians of vacuum expectation values. Since the Pfaffian of an anti-symmetric matrix is the square root of the determinant of this matrix, these BKP tau-functions are square roots of certain determinants and in fact Date, Jimbo, Kashiwara and Miwa show that it is the square root of a certain KP tau-function, provided one puts the even KP times equal to zero. This fact, was used by the author and A. Yu. Orlov in \cite{pfaff},  in another realization of this $b_\infty$ group orbit, to give a representation theoretical proof of the fact  that Pfaff Lattice tau-functions are square roots of 2D Toda lattice tau-functions. 
J. Harnad and A.Yu. Orlov \cite{HO1}-\cite{HO3} also use this observation to 
express  KP and BKP tau-functions as sums over products of pairs of $Q$ Schur  functions. 

In  \cite {KvdLmodKP}, \cite{KvdLB2},  V.G.   Kac  and the author gave explicit formulas for all KP and also BKP tau-function. In this paper,  we calculate for every polynomial BKP tau-function its corresponding square, i.e. the corresponding KP tau-function. 

In section \ref{S7} we give a representation theoretical explanation, at least in the polynomial case,  of a recent result of A. Alexandrov \cite{Alex},  viz. that a KdV tau-function, which is a KP tau-function that does not depend on the even times,  becomes a BKP tau-function when one divides all KdV times by 2.

\section{The  fermionic formulation of KP}
Consider the infinite  matrix group
$GL_{\infty}$, consisting of all complex matrices  $G= (g_{ij})_{i,j \in {\mathbb Z}}$ which are 
 invertible and all but a finite number of $g_{ij} -
\delta_{ij}$ are $0$. We denote its Lie algebra by $gl_{\infty}$  consisting of all complex matrices  $g= (g_{ij})_{i,j \in {\mathbb Z}}$ for which are  all but a finite number of $g_{ij}$ are $0$. Both  the group and its Lie algebra 
act  naturally on the vector space 
${\mathbb C}^{\infty} = \bigoplus_{j \in {\mathbb Z}} {\mathbb
C} e_{j}$  (via the usual formula
$E_{ij} (e_{k}) = \delta_{jk} e_{i}$).

The semi-infinite wedge representation \cite{KPeterson}, \cite {KvdLmodKP}.
 $F =
\Lambda^{\frac{1}{2}\infty} {\mathbb C}^{\infty}$ is the vector space
with a basis consisting of all semi-infinite monomials of the form
$e_{i_{0}} \wedge e_{i_{1}} \wedge e_{i_{2}} \ldots$, where $i_{0} >
i_{1} > i_{3} > \ldots$ and $i_{\ell +1} = i_{\ell} -1$ for $\ell >>
0$.  One defines the  representation $R$ of $GL_{\infty}$ and $r$
of $gl_{\infty}$ on $F$ by
$$
R(G) (e_{i_{1}} \wedge e_{i_{2}} \wedge e_{i_{3}} \wedge \cdots) = G
e_{i_{1}} \wedge G e_{i_{2}} \wedge Ge_{i_{3}} \wedge \cdots .
$$

The corresponding representation $r$ of the Lie algebra
$g\ell_{\infty}$
of $GL_\infty$can be described
in  terms of a Clifford algebra.  Define the wedging and contracting
operators $\psi^{+}_{j}$ and $\psi^{-}_{j}\ \ (j \in {\mathbb Z} +
\frac{1}{2})$ on $F$ by
$$\begin{aligned}
&\psi^{+}_{j} (e_{i_{0}} \wedge e_{i_{1}} \wedge \cdots ) =  
e_{-j+\frac12}\wedge e_{i_{0}} \wedge e_{i_{1}} \cdots, \\
&\
\psi^{-}_{j} (e_{i_{0}} \wedge e_{i_{1}} \wedge \cdots ) = \begin{cases} 0
&\text{if}\ j+\frac12  \neq i_{s}\ \text{for all}\ s \\
(-1)^{s} e_{i_{0}} \wedge e_{i_{1}} \wedge \cdots \wedge
e_{i_{s-1}} \wedge e_{i_{s+1}} \wedge \cdots &\text{if}\ j+\frac12 = i_{s}.
\end{cases}
\end{aligned}
$$
These operators satisfy the relations
$(i,j \in {\mathbb Z}+\frac{1}{2}, \lambda ,\mu = +,-)$:
\begin{equation}
\label{rel1}
\psi^{\lambda}_{i} \psi^{\mu}_{j} + \psi^{\mu}_{j}
\psi^{\lambda}_{i} = \delta_{\lambda ,-\mu} \delta_{i,-j}, 
\end{equation}
hence they generate a Clifford algebra, which we denote by ${\cal C}\ell$.
Introduce the following elements of $F$ $(m \in {\mathbb Z})$:
$$|m\rangle = e_{m } \wedge e_{m-1 } \wedge
e_{m-2 } \wedge \cdots .$$
It is clear that $F$ is an irreducible ${\cal C}\ell$-module such that
$$\psi^{\pm}_{j} |0\rangle = 0, \ \text{for}\ j > 0 . $$

It will be convenient to define also the oposite spin module with vacuum vector $\langle 0 |$, here
\[
\langle 0 |\psi^\pm_j=0,\ \text{for}\ j <0,
\]
and for $m>0$ one defines
\[
\langle \pm m|=\langle 0|\psi^\mp_{\frac12}\psi^\mp_{\frac32}\cdots\psi^\mp_{m-\frac12}.
\]
The vacuum expectation value is defined as $\langle a\rangle =\langle 0| a| 0\rangle$ and $\langle 0| 1| 0\rangle=1$.
It is straightforward that the representation $r$ of $g\ell_\infty$ is given by the
 formula
$r(E_{ij}) = \psi^{+}_{-i+\frac12} \psi^{-}_{j-\frac12 }. $
Define the {\it charge decomposition}
$$F = \bigoplus_{m \in {\mathbb Z}} F^{(m)}, \quad \text{
where }
\text{charge}(|m\rangle ) = m\ \text{and charge} (w^{\pm}_{j}) =
\pm 1. $$
The space $F^{(m)}$ is an irreducible highest weight
$g\ell_{\infty}$-module, where 
$|m\rangle$ is its highest weight vector, i.e.
$$
r(E_{ij})|m\rangle = 0 \ \text{for}\ i < j,\quad
r(E_{ii})|m\rangle = 0\  (\text{resp.}\ = |m\rangle ) \ \text{if}\ i > m\
(\text{resp. if}\ i < m).
$$
Let $S$ be the following operator on $F\otimes F$
\[
S=\sum_{i\in\mathbb{Z}+
\frac{1}{2}} \psi_i^+ \otimes \psi_{-i}^-
\]
and let 
$${\cal O}_m
= R(GL_{\infty})|m\rangle \subset F^{(m)}$$ be the $GL_{\infty}$-orbit
of the highest weight vector $|m\rangle$.

\begin{theorem}{\rm  (\cite{KPeterson}, Theorem 5.1)}
Let $M$ be an integer and  let $f=\oplus_{m\in \mathbb{Z}}f_m\in\oplus_{m\in \mathbb{Z}}F^{(m)}$ be such that all $f_m\not = 0$ and $f_m=|m\rangle$ for $m<M$.
Then $f\in \oplus_{m\in \mathbb{Z}} {\cal O}_m$ if and only if for all $k,\ell\in \mathbb{Z}$, such that $k\ge \ell$, one has
\begin{equation}\
\label{modKP}
S(f_k\otimes f_\ell)=
\sum_{i\in\mathbb{Z}+
\frac{1}{2}} \psi_i^+ f_k\otimes \psi_{-i}^- f_\ell =0\, .
\end{equation}
\end{theorem}

Equation (\ref{modKP}) is called the $(k-\ell)$-th modified KP hierarchy in the fermionic picture. The 0-th modified KP is the KP hierarchy.
The collection of all such equations $k,\ell\in\mathbb{Z}$ with $k\ge \ell$ is called the (full) MKP hierachy in the fermionic picture.

\section{The fermionic formulation of BKP}

The Lie group $B_\infty$ and the  corresponding Lie algebra $b_\infty$  can be defined using the following bilinear form on  $\mathbb{C}_\infty$, see e.g. \cite{Kacbook}, section 7.11:
\begin{equation}
\label{bilinform}
(e_i,e_j)_B=(-1)^i\delta_{i,-j}.
\end{equation}
Then 
\[
\begin{aligned}
B_\infty&=\{ G\in GL_\infty | (G(v),G(w))_B=(v,w)_B \ \mbox{for all } v,w\in \mathbb{C}^\infty\},\\
b_\infty&=\{ g\in gl_\infty | (g(v),w)_B+(v,g(w))_B=0 \ \mbox{for all } v,w\in \mathbb{C}^\infty\},\\
\end{aligned}
\]
The elements $F_{jk}=E_{-j,k}-(-1)^{j+k}E_{-k,j}=-(-1)^{j+k}F_{kj}$, with $j>k$
 form a basis of $\oi$.
Note that 
\[
\begin{aligned}
r(F_{jk})& =\psi^+_{j+\frac12}\psi^-_{k-\frac12}-(-1)^{j+k} \psi^+_{k+\frac12}\psi^-_{j-\frac12}
.
\end{aligned}
\]
This suggests to define  linear anti-involutions on the Clifford algebra $C\ell$, which respects the relations \eqref{rel1}:
\begin{equation}
\label{anti}
\begin{aligned}
\iota_B(\psi^+_{j+\frac12})&=(-1)^j \psi^-_{j-\frac12},\quad&\iota_B( \psi^-_{k-\frac12})=(-1)^k \psi^+_{k+\frac12}.
\end{aligned}
\end{equation}
This induces via $r$ the following the following anti-involution on $gl_\infty$
$$\iota_B(E_{jk})=(-1)^{j+k}E_{-k,-j}$$
thus
$$\oi=\{g\in\gli |\iota_B(g)=-g\}.$$

Instead of $\psi_i^\pm$,  and inspired by \cite{DJKM3} (see also \cite{You}), we choose different operators that generate $C\ell$, viz.  eigenvectors of $\iota_B$ 
\begin{equation}
\label{phi-hat-phi}
\begin{aligned}
\phi_i&= \frac{\psi_{i+\frac12}^++(-1)^i\psi_{i-\frac12}^-}{\sqrt 2},\quad  &\hat\phi_i&= \sqrt {-1}\frac{\psi_{i+\frac12}^+-(-1)^i\psi_{i-\frac12}^-}{\sqrt 2}, \ \mbox{for } i\in \mathbb{Z} ,
\end{aligned}
\end{equation}
related to $b_\infty$.
These elements satisfy the following relations:
\begin{equation}
\begin{split}
\phi_i\phi_j+\phi_j\phi_i=(-1)^i \delta_{i,-j},\  \phi_i\hat\phi_j+\hat\phi_j\phi_i= 0, \ \hat\phi_i\hat\phi_j+\hat\phi_j\hat\phi_i= (-1)^i \delta_{i,-j}, \ i,j\in\mathbb{Z}.
\end{split}
\end{equation} 
Thus, we have the following symmetric bilinear form 
\begin{equation}
\label{1}
\begin{split}
(\phi_i,\phi_j)_B=(\hat\phi_i,\hat\phi_j)_B=(-1)^i\delta_{i,-j},\ (\hat \phi_i,\phi_j)_B=0,\ \mbox{for }i,j\in\mathbb{Z} .
\end{split}
\end{equation}
We observe that
\begin{equation}
\label{rphi}
\begin{aligned}
r(F_{jk})& = \frac{(-1)^k}2(\phi_j\phi_k-\phi_k\phi_j)+\frac{(-1)^k}2(\hat\phi_j\hat\phi_k-\hat\phi_k\hat\phi_j), \ \mbox{for } i,j\in\mathbb{Z},
\end{aligned}
\end{equation}
and that in both cases 
\[
\phi_j|0\rangle=\hat\phi_j|0\rangle=0, \ \mbox{for }j>0.
\]
The action of $\phi_0$ and $\hat\phi_0$ is special and one has  
\begin{equation}
\label{1_B}
\begin{split}
&\phi_0|0\rangle =\frac{1}{\sqrt 2}|1_B\rangle :=\frac1{\sqrt 2}|-1\rangle,\ \hat \phi_0|0\rangle= -\frac{\sqrt{-1}}{\sqrt 2}|\hat 1_B\rangle :=-\frac{\sqrt{-1}}{\sqrt 2}|-1\rangle, \\
&\phi_0|-1\rangle =\frac1{\sqrt 2}|0\rangle,\ \hat \phi_0|-1\rangle=\frac{\sqrt{-1}}{\sqrt 2}|0\rangle
\end{split}
\end{equation}
and 
\begin{equation}
\label{1_B2}
\begin{split}
&\langle 0| \phi_0 =\frac1{\sqrt 2}\langle 1_B|:=\frac1{\sqrt 2}\langle -1|,\ \langle 0|\hat \phi_0 =\frac{\sqrt{-1}}{\sqrt 2}\langle -1|:=\frac{\sqrt{-1}}{\sqrt 2}\langle \hat 1_B|, \\
&
\langle -1| \phi_0 =\frac1{\sqrt 2}\langle 0|,\ \langle -1|\hat \phi_0 =-\frac{\sqrt{-1}}{\sqrt 2}\langle 0|,
\end{split}
\end{equation}
which gives that  $ |1_B\rangle= |\hat 1_B\rangle$,   $\langle 1_B|=\langle \hat 1_B|$ and  
\[
\langle 0|\hat \phi_0 \phi_0|0\rangle=-\langle 0| \phi_0\hat \phi_0|0\rangle=
\langle -1| \phi_0 \hat\phi_0|-1\rangle=-\langle -1| \hat\phi_0 \phi_0|-1\rangle=
\frac{\sqrt{-1}}2.
\]
Note that 
\begin{equation}
\label{psiphi}
\begin{aligned}
\psi_{i+\frac12}^+
&= \frac{\phi_{i}-\sqrt{-1}\hat\phi_i}{\sqrt 2}, \quad \
\psi_{i+\frac12}^-
&=(-1)^i\frac{\phi_{i}+\sqrt{-1}\hat\phi_i}{\sqrt 2},\ \mbox{for } i\in \mathbb{Z} ,
\end{aligned}
\end{equation}

The $gl_\infty$ level one representation  $r$, when restricted to $b_\infty$ gives a level two representation of this orthogonal infinite dimensional Lie algebra. The formula's \eqref{rphi} make it possible to define the level one spin representations of this algebra in two ways on $F_B$, $\hat F_B$:
\begin{equation}
\label{rBD}
\begin{split}
r_B(F_{jk})=\frac{(-1)^k}{2}(\phi_j\phi_k-\phi_k\phi_j)\ \mbox{or}\\
\hat r_B(F_{jk})=\frac{(-1)^k}{ 2}(\hat\phi_j\hat\phi_k-\hat\phi_k\hat \phi_j) .
\end{split}
\end{equation}
Each module splits in to two irreducible level one representations  
$F_B=F_B^0\oplus F_B^1$, $\hat F_B=\hat F_B^0\oplus\hat F_B^1$ 
for  $r_B$, $\hat r_B$, respectively, with highest weight vectors $|0\rangle$ and   $|1_B\rangle$. 

The elements $\phi_{j_1}\phi_{j_2}\cdots\phi_{j_p}|0\rangle$ ( resp. $\hat\phi_{j_1}\hat\phi_{j_2}\cdots\hat \phi_{j_p}|0\rangle$)
with $j_1<j_2<\cdots <j_p\le 0$ form a basis of $F_B$ (resp. $\hat F_B$).

Let $S_B$, $\hat S_B$   be the following operator on $F_B\otimes F_B$,  $\hat F_B\otimes \hat F_B$, respectively:
\[
\label{S}
\begin{split}
&S_B=\sum_{j\in\mathbb{Z}} (-1)^j \phi_j\otimes \phi_{-j},\qquad 
\hat S_B=\sum_{j\in\mathbb{Z}} (-1)^j \hat\phi_j\otimes\hat \phi_{-j}.
\end{split}
\]

To define the hierarchies in the B case, we assume that 
 $\tau \in F_B^\nu$ (resp.  $\tau \in\hat F_B^\nu$),  has the form $\tau=g|\nu \rangle$, it is called a {\it tau-function} of the {\it BKP} hierarchy if,
 \begin{equation}
      \label{BKP}
      S_B(g|\nu \rangle \otimes g|\nu \rangle) = g\phi_0|\nu \rangle \otimes  g\phi_0|\nu\rangle  \qquad ( \mbox{resp. } \hat S_B(g|\nu \rangle  \otimes g|\nu \rangle ) =g\hat \phi_0|\nu \rangle  \otimes  g\hat\phi_0|\nu \rangle ).
    \end{equation}
In fact, see e.g.  \cite{KvdLB} or \cite{KvdLB2}, equation \eqref{BKP} describes the $B_\infty$-orbit of $|\nu\rangle$, where $\nu=0$ or $1$.
\section{Vertex operators }

\noindent
In this section, we want to realize the spin module $F$ in two different ways. An indication  that these isomorphisms exist is given by the following gradation of our module $F$. Define
\[
	\deg (|0 \rangle)=0,\  \deg( \phi_{-i})= \deg(\hat \phi_{-i})=\deg(\psi^\pm_{-i\pm\frac12})=i
\]
and let $F_k=\{ f\in F | \deg(f)=k\}$. The character formule $\dim_q F=\sum_{k\in\mathbb{Z}} \dim(F_k) q^k$, is clearly equal to
\[
2\prod_{k=1}^\infty (1+q^k)^2,
\]
since the elements $\psi^+_{i_1}\psi^+_{i_2}\cdots\psi^+_{i_m}\psi^-_{j_1}\psi^-_{j_2}\cdots\psi^-_{j_n}|0\rangle$, with
$i_1<i_2<\cdots< i_m<0$ and $j_1<j_2<\cdots< j_n<0$, form a basis of $F$.
We can rewrite this character formula in two different ways. The first one is 
\[
2\prod_{k=1}^\infty (1+q^k)^2 =2\prod_{k=1}^\infty \left ( (1+q^k) \frac{1-q^k}{1-q^k}\right)^2=
2\prod_{k=1}^\infty \left (  \frac{1-q^{2k}}{1-q^k}\right)^2=2\prod_{k=1}^\infty \left (  \frac{1}{1-q^{2k-1}}\right)^2
\]
and for the second one we use the Jacobi triple product identity wich gives that 
\[
2\prod_{k=1}^\infty (1+q^k)^2 =\sum_{j\in \mathbb{Z} }q^{\frac{j(j-1)}2}\prod_{k=1}^\infty   \frac{1}{1-q^{k}}.
\]

We define two  isomorphisms $\sigma$ and $\overline \sigma$, such that $\sigma(F)=B$ and $\overline\sigma= \overline B$, where 
\begin{equation}
\label{Bspace}
B=\mathbb{C} [q, q^{-1}, t_k |  k=1,2,\ldots],\qquad \overline B=\mathbb{C} [\theta, \overline t_k, \hat t_k | k=1,3,5,\ldots].
\end{equation}
Here $\theta$ is a Grassmann variable, i.e. $\theta^2=0$, which commutes with all the other indeterminates.

The isomorphisms  are uniquely determined by the following properties  \cite{KvdLB}. First, $\sigma(|0\rangle)=\overline\sigma (|0\rangle)=1$. Second,
\begin{equation}
\label{charged-vertex}
\sigma \psi^{\pm }(z)\sigma^{-1}=\sum_{k\in\frac12+\mathbb{Z}}\sigma \psi^{\pm }_k\sigma^{-1}z^{-k-\frac12}=
q^{\pm 1}z^{\pm q\frac{\partial}{\partial q}}
\exp\left(\pm \sum_{i=1}^\infty t_iz^i\right)
\exp\left(\mp \sum_{i=1}^\infty \frac{\partial}{\partial  t_i}\frac{z^i}{i}\right)
\end{equation}
and
\begin{equation}
\label{neutral-vertex}
\begin{aligned}
\overline \sigma\phi(z)\overline\sigma^{-1}&=\sum_{k\in\mathbb{Z}}\overline\sigma \phi_k\overline\sigma^{-1}z^{-k}=\frac{\theta +\frac{\partial}{\partial \theta}}{\sqrt 2} 
\exp\left( \sum_{i>0,\,  odd}^\infty\overline t_iz^i\right)
\exp\left(-2 \sum_{i>0,\, odd}^\infty \frac{\partial}{\partial \overline t_i}\frac{z^i}{i}\right),\\
\overline \sigma\hat\phi(z)\overline\sigma^{-1}&=\sum_{k\in\mathbb{Z}}\overline\sigma \hat\phi_k\overline\sigma^{-1}z^{-k}=\sqrt{-1}\frac{-\theta +\frac{\partial}{\partial \theta}}{\sqrt 2} 
\exp\left( \sum_{i>0,\,  odd}^\infty\hat t_iz^i\right)
\exp\left(-2 \sum_{i>0,\, odd}^\infty \frac{\partial}{\partial \hat t_i}\frac{z^i}{i}\right).
\end{aligned}
\end{equation}
Note that 
\[
\sigma(|m\rangle)= q^m,\quad\mbox{and } \overline \sigma (|-1\rangle)=\theta.
\]
Both isomorphisms make it possible to express an element $f=g|0\rangle \in F$ as  function in $B$ or $\overline B$, viz.
\[
\sigma(f)=\sum_{k\in\mathbb{Z}} g_k(t)q^k,\quad \overline \sigma(f)= \overline g_0(\overline t,\hat t)+\overline g_1(\overline t,\hat t)\theta.
\]

To determine these functions it will  be convenient to introduce the oscillator algebra associated to the  above fermionic fields. Let $:ab:=ab-\langle0| ab|0 \rangle$ stand for the normal ordered product of two elements. Define
\begin{equation}
\label{charged-alpha}
\alpha(z)=\sum_{k\in\mathbb{Z}}\alpha_kz^{-k-1}=:\psi^+(z)\psi^-(z):\, ,
\end{equation}
and 
\begin{equation}
\label{neutral-alpha}\beta(z) = \sum\limits_{k\in 2 \mathbb{Z}+1} \beta_k
 z^{-k-1} = :
\phi(z)  \frac{\phi(-z)}{z}:, \qquad
\hat\beta(z) = \sum\limits_{k\in 2 \mathbb{Z}+1} \hat\beta_k
 z^{-k-1} = :
\hat\phi(z)  \frac{\hat\phi(-z)}{z}:,
\end{equation}
then 
\[
\sigma(\alpha(z))=q\frac{\partial}{\partial q}z^{-1}+\sum_{k=1}^\infty 
\left( kt_k z^{k-1}+\frac{\partial}{\partial t_k} z^{-k-1} \right)
\, 
\]
and 
\[
\begin{split}
\overline\sigma(\beta(z))&=\sum\limits_{0<k\in 2 \mathbb{Z}+1}\left(k\overline t_k z^{k-1}+2\frac{\partial}{\partial \overline t_k }z^{-k-1 }    
 \right),
\\
\overline\sigma(\hat\beta(z))&=\sum\limits_{0<k\in 2 \mathbb{Z}+1}\left(k\hat t_k z^{k-1}+2\frac{\partial}{\partial \hat t_k }z^{-k-1 }    
 \right).
\end{split}
\]
We observe that
\begin{equation}
\label{alphabeta}
\begin{aligned}
\beta(z)+\hat\beta(z)&= :
\phi(z)  \frac{\phi(-z)}{z}:+ :
\hat\phi(z)  \frac{\hat\phi(-z)}{z}:\\
&=
\sum_{i,j\in\mathbb{Z}} {z^{-i-j-1}}
\left(:\psi_{i+\frac12}^+\psi_{j-\frac12}^-:
+(-1)^{i+j}:\psi_{i-\frac12}^-\psi_{j+\frac12}^+:\right)
\\
&=:\psi^+(z)\psi^-(z):+:\psi^+(-z)\psi^-(-z):\\
&=\alpha(z)+\alpha(-z)\\
&=2\sum\limits_{k\in 2 \mathbb{Z}+1}\alpha_kz^{-k-1}.
\end{aligned}
\end{equation}
Define 
\begin{equation}
\label{H}
H(s)=  \sum_{k>0}s_k  \alpha_k,\quad \overline H(s)
=\sum_{k>0,{\rm odd}}\frac{s_k}2 \beta_k,\quad \mbox{and }
 \hat H(s)
=\sum_{k>0,{\rm odd}}\frac{s_k}2 \hat\beta_k,
\end{equation} 
then  
\[
\sigma(H(s))= \sum_{k>0}{s_k}  \frac{\partial}{\partial t_k}, \quad
\overline\sigma(\overline H(s))= \sum_{k>0,{\rm odd}}{s_k} \frac{\partial}{\partial \overline t_k},\quad
\overline\sigma(\hat H(s))= \sum_{k>0,{\rm odd}}{s_k} \frac{\partial}{\partial \hat t_k}.
\]
One has
\[
H(s_1,0,s_3,0,s_5,0,\ldots)=\overline H(s)+\hat H(s)
\]
and
\begin{lemma}
\label{lemma1} (a)
\[
\exp (H(s)) | 0\rangle=\exp (\overline H(s)) | 0\rangle=\exp (\hat H(s)) | 0\rangle= | 0\rangle,
\]
(b) 
\[
\exp(H(s))\psi^{\pm }(z)\exp(-H(s))= \psi^{\pm }(z)\exp\left(\pm \sum_{k>0} s_k z^k\right).
\]
(c)
\[
\begin{aligned}
\exp(\overline H(s))\phi(z)\exp(-\overline H(s))=& \phi (z)\exp\left(\sum_{k>0, odd}s_k z^k\right),\\
\exp(\hat H(s))\hat \phi(z)\exp(-\hat H(s))=& \hat\phi (z)\exp\left(\sum_{k>0, odd}s_k z^k\right),
\end{aligned}
\]
and 
\[
\begin{aligned}
\exp(\overline H(s))\hat\phi(z)\exp(-\overline H(s))=&\hat \phi (z),\\
\exp(\hat H(s))  \phi(z)\exp(-\hat H(s))=& \phi (z) .
\end{aligned}
\]
\end{lemma}
{\bf Proof.} (a) follows from the fact that all $\alpha_k|0\rangle=\beta_k|0\rangle=\hat\beta_k|0\rangle= 0$ for all $k>0$.\\
(b) (resp.  (c)) follows from  the fact that 
 $[ \alpha_{k}^{i},\psi^{\pm j}(z)]=\pm\delta_{ij} z^k\psi^{\pm j}(z)$ and  $[\frac12 \beta_{k},\phi(z)]=z^k\phi(z)$ (resp.
 $[\frac12 \hat \beta_{k},\hat\phi(z)]=z^k\hat\phi(z)$).
\hfill$\square$
\\
\
\\
We deduce from \eqref{alphabeta} and part (b) and (c) of the above lemma that
\begin{equation}
\label{alphabeta2}
\begin{aligned}
\exp(\overline H(s)+\hat H(s))\psi^\pm(z)\exp(-\overline H(s)-\hat H(s))=&
\exp(H(s)\psi^{\pm }(z)\exp(-H(s)|_{all \ s_{2k}=0}\\
=& \psi^{\pm }(z)\exp\left(\pm \sum_{k>0, odd}s_k z^k\right).
\end{aligned}
\end{equation}

Now let $\exp (H(s))$ act on $f= g|0\rangle\in F$. Since we can decompose such an element as $f=\sum_{k\in\mathbb{Z}}f_k$, where each $f_k\in F^{(k)}$, thus we can write 
\[
f= g|0\rangle
=\sigma^{-1}\left(\sum_k g_k(t)q^k\right)
=\sum_k\sigma^{-1}\left( g_k(t)\right)]k\rangle .
\]
This  gives
\[
\exp(H(s))f=\exp(H(s))g|0\rangle = \sum_k \sigma^{-1}\left(g_k(t+s)\right)|k\rangle.
\]
Now, let $T_k(s)$ be the coefficient of the highest weight vector $|k\rangle$ in the above expression, then 
\[
T_k(s)=\langle k|\exp (H(s))f=g_k(s).
\]
Thus 
\begin{equation}
\label{expH}
\sigma(f)=
\sigma(g|0\rangle)=\sum_{k\in\mathbb{Z}}
\langle k| \exp(H(t))g|0\rangle q^k.
\end{equation}
In a similar way one obtains:
\begin{equation}
\label{expH2}
\overline\sigma(f)=
\overline\sigma(g|0\rangle)=
\left(\langle 0| +\theta \langle -1|\right)\exp\left(\overline H(\overline t)+\hat H(\hat t)\right)g|0 \rangle .
\end{equation}

\section{The  bosonic formulation of MKP and BKP}

Under the isomorphism $\sigma$  we can rewrite (\ref{modKP}), using (\ref{charged-vertex}), to obtain  the
{\bf  MKP hierarchy:}
\\
{\it 
Let 
$[z]=(z,\frac{z^2}2,\frac{z^3}3,\ldots)$, $y=(y_1,y_2,\ldots)$, and ${\rm Res}\,  \sum_i f_iz^i dz=f_{-1}$, then 
\begin{equation}
\label{modKP2}
{\rm Res}\,   z^{k-\ell}\tau_k(t-[z^{-1}])\tau_\ell(y+[z^{-1}]) \exp\left(\sum_{i=1}^\infty (t_i-y_i)z^i\right)dz=0,\qquad k\ge \ell.
\end{equation}
}The equations (\ref{modKP2})
first appeared in \cite{JM}, $(2.4)_{l,l'}$.
 In a similar way, but now using the isomorphism $\overline \sigma$ and \eqref{neutral-vertex} we can reformulate \eqref{BKP}, to obtain the {\bf  BKP hierarchy} \cite{DJKM3},\cite{JM},\cite{KvdLB}:
\\
{\it 
Let 
$[z]_{odd}=(z,\frac{z^3}3,\ldots)$, $y=(y_1,y_3\ldots)$, then 
\begin{equation}
\label{BKP2}
{\rm Res}\,   \tau(t-2[z^{-1}]_{odd})\tau(y+2[z^{-1}]_{odd}) \exp\left(\sum_{i=1}^\infty (t_{2i-1}-y_{2i-1})z^{2i-1}\right)\frac{dz}z=\tau(t)\tau(y).
\end{equation}
}

\section{Polynomial tau-functions}
A polynomial tau-function of the  BKP  hierarchy (\ref{BKP}) corresponds to an element (cf. \cite{KvdLB2})
\begin{equation}
\label{tauk}
\begin{split}
f^{k}=v_1v_{2}\cdots v_k|0\rangle,\quad\mbox{with } v_i=\sum_{ j\in\mathbb{Z}}(-1)^j  (v_i,\phi_{-j})\phi_j\in\mathbb{C}^\infty,\ \mbox{or }\\
\hat f^{k}=\hat v_1 \hat v_{2}\cdots \hat v_k|0\rangle,\quad\mbox{with } \hat v_i=\sum_ { j\in\mathbb{Z}}(-1)^j  (\hat v_i,\hat\phi_{-j})\hat\phi_j\in\hat{\mathbb{C}}^\infty, 
\end{split}
\end{equation}
This is obvious from the fact that $v_i\otimes v_i$ commutes with $S_B$. Indeed,
 using  that an element $v\in \mathbb{C}^\infty$ can be written as $v=\sum_ j(-1)^j  (v,\phi_{-j})\phi_j$,
$(v,v)=\sum_ j(-1)^j  (v,\phi_{-j})(v,\phi_j)$ and that $v^2=\frac{(v,v)}2$, one finds
\[
\begin{aligned}
(v\otimes v )S_B&=\sum_j (-1)^j v\phi_j\otimes v\phi_{-j}\\
&=\sum_j ((v,\phi_j)-\phi_j v)\otimes((-1)^j (v,\phi_{-j})-(-1)^j\phi_{-j}v)\\
&=(v,v)1\otimes 1-v^2\otimes 1-1\otimes v^2+\sum_j (-1)^j \phi_jv\otimes \phi_{-j}v\\
&=0+ \sum_j (-1)^j \phi_jv\otimes \phi_{-j}v\\
&=S_B (v\otimes v).
\end{aligned}
\]
Thus if $f^{k-1}$ satisfies \eqref{BKP} then $vf^{k-1}$, again satisfies \eqref{BKP}.

In order to express BKP  tau-functions as the square root of a certain KP tau-function, as was shown  in \cite{DJKM3}, we want to calculate 
\begin{equation}
\label{gk}
g^k=v_1v_{2}\cdots v_k \hat v_1 \hat v_{2}\cdots \hat v_k|0\rangle,\ \mbox{with } v_i=\sum_{j} a_{ij}\phi_j,\ \hat v_i=\sum_{j} a_{ij}\hat\phi_j,
\end{equation}
where for every $1\le i\le k$ the coefficients $a_{ij}$ that appear in $v_i$ and $\hat v_i$ are equal.
Now, 
\[
g^k=(-1)^{\frac{k(k-1)}2 }v_1\hat v_1v_{2}\hat v_{2}\cdots v_k  \hat v_k|0\rangle\]
and 
\[
\begin{split}
v_i\hat v_i=&\sum_{j}a_{ij}\phi_j\sum_{\ell}a_{i\ell}\hat \phi_\ell\\
=& \sqrt {-1}\sum_{j,\ell}a_{ij} a_{i\ell} \frac{\psi_{j+\frac12}^++(-1)^j\psi_{j-\frac12}^-}{\sqrt 2}
\frac{\psi_{\ell+\frac12}^+-(-1)^\ell\psi_{\ell-\frac12}^-}{\sqrt 2}\\
=& -\frac{\sqrt {-1}}2\sum_{j,\ell}a_{ij} a_{i\ell} \left((-1)^\ell \psi_{j+\frac12}^+\psi_{\ell-\frac12}^-
-(-1)^j \psi_{j-\frac12}^-
\psi_{\ell+\frac12}^+\right).
\end{split}
\]
Hence the element $g^k\in F^{(0)}$. Moreover, 
\begin{proposition}
The element $g^k\in F^{(0)}$ of \eqref{gk}  satisfies the KP hierarchy \eqref{modKP}. 
\end{proposition}
{\bf Proof.}
To prove this, it  will be sufficient to show tha $v_i\hat v_i\otimes v_i\hat v_i$ commutes with $S$. Note that up to a constant  the element 
$v_i\hat v_i$
is of the form $w^+v^--v^-w^+$ where 
\[
w^+=\sum_j a_{ij}\psi^+_{j+\frac12}, \qquad v^-=\sum_j (-1)^j a_{ij}\psi^-_{j-\frac12}.
\]
Now, observe that $v^-v^-=w^+w^+=0$, thus
\[
\begin{aligned}
S&(w^+v^--v^-w^+)\otimes(w^+v^--v^-w^+)\\
&=\sum_k \psi^+_k(w^+v^--v^-w^+)\otimes\psi_{-k}^-(w^+v^--v^-w^+)\\
&=
\sum_k \left(-2( \psi^+_k,v^-)w^++(w^+v^--v^-w^+) \psi^+_k\right)\otimes
\left(2( \psi_{-k}^-,w^+)v^- +(w^+v^--v^-w^+) \psi_{-k}^-\right)\\
&=-4(w^+,v^-) w^+\otimes v^-
-2 w^+\otimes (w^+v^--v^-w^+) v^-
+2(w^+v^--v^-w^+) w^+\otimes v^-\\
&\qquad\qquad +(w^+v^--v^-w^+)\otimes(w^+v^--v^-w^+)S\\
&=-4(w^+,v^-) w^+\otimes v^-
+2w^+\otimes  (w^+,v^-) v^-
+2 (w^+,v^-) w^+\otimes  v^-\\
&\qquad\qquad +(w^+v^--v^-w^+)\otimes(w^+v^--v^-w^+)S\\
&=(w^+v^--v^-w^+)\otimes(w^+v^--v^-w^+)S\\
\end{aligned}
\]
Thus $v_i\hat v_i\otimes v_i\hat v_i$ commutes  with $S$.
\hfill$\square$
\\
\ 
\\
In order to calculate the corresponding tau-functions, we first calculate several vacuum expextation values. 
\[
\begin{aligned}
\exp&\left( \overline H(\overline t)+\hat H(\hat t)\right)v_i\exp\left(- \overline H(\overline t)-\hat H(\hat t)\right)=
 \exp\left( \overline H(\overline t)+\hat H(\hat t)\right)\sum_{j\ge -N_i}a_{ij}\phi_j\exp\left(- \overline H(\overline t)-\hat H(\hat t)\right)\\
&={\rm Res} \sum_{j\ge -N_i}a_{ij}z^{j}          \exp\left( \overline H(\overline t)+\hat H(\hat t)\right)\phi(z)\exp\left(- \overline H(\overline t)-\hat H(\hat t)\right) \frac{dz}z\\
&={\rm Res}\,  z^{-N_i-1}\sum_{j\ge 0}a_{i,j-N_i}z^j \exp\left(\sum_{k>0, odd}\overline t_k z^k\right)\phi(z)\frac{dz}z.
\end{aligned}
\]
Here we assume that $a_{i,-N_i}\ne 0$.
We then write 
\begin{equation}
\label{ac}
\sum_{j\ge 0}a_{i,j-N_i}z^j =a_{i,-N_i} \exp \left( \sum_{i=1}^\infty c_{ij}z^j\right).
\end{equation}
Hence  for $\overline t=(\overline t_1,0,\overline t_3,0,\ldots)$, we find that
\[
\exp\left( \overline H(\overline t)+\hat H(\hat t)\right)v_i\exp\left(- \overline H(\overline t)-\hat H(\hat t)\right)=\\
a_{i,-N_i}{\rm Res}\,  z^{-N_i} \exp\left(\sum_{k>0}(\overline t_k+ c_{ik}) z^k\right)\phi(z)\frac{dz}z
\]
Thus,
\[
\begin{aligned}
&\langle 0|\exp\left( \overline H(\overline t)+\hat H(\hat t)\right)v_iv_j\exp\left(- \overline H(\overline t)-\hat H(\hat t)\right)|0\rangle =\\
&a_{i,-N_i}a_{j,-N_j}{\rm Res}\,  z^{-N_i}w^{-N_j} \exp\left(\sum_{k>0}(\overline t_k+ c_{ik}) z^k\right)
 \exp\left(\sum_{k>0}(\overline t_k+ c_{jk}) w^k\right)
\langle 0|\phi(z)\phi(w)|0\rangle \frac{dz}z\frac{dw}w.
\end{aligned}
\]
Using that
\[
\langle 0|\phi(z)\phi(w)|0\rangle=  (zw)^{-1}\left( \frac12+ \sum_{i=1}^\infty \left( -\frac{w}{z}\right)^i\right)
\]
and 
\[
\exp\left(\sum_{k>0} t_kz^k\right)=\sum_{j=0}^\infty s_j(t)z^j,
\]
we find that 
\[
\langle 0|\exp\left( \overline H(\overline t)+\hat H(\hat t)\right)v_iv_j\exp\left(- \overline H(\overline t)-\hat H(\hat t)\right)|0\rangle =
a_{i,-N_i}a_{j,-N_j}\overline\chi_{N_i,N_j}(\overline t+c_i, \overline t +c_j),
\]
where 
\begin{equation}
\label{chi}
\overline \chi_{N,M}(s,t)=\frac 12 s_N(t)s_M(s)+\sum_{k=1}^M (-1)^k s_{N+k}(t)s_{M-k}(s).
\end{equation}
Clearly,
\[
\langle 0|\exp\left( \overline H(\overline t)+\hat H(\hat t)\right)\hat v_i\hat v_j\exp\left(- \overline H(\overline t)-\hat H(\hat t)\right)|0\rangle =
a_{i,-N_i}a_{j,-N_j}\overline \chi_{N_i,N_j}(\hat t+c_i, \hat t +c_j)
\]
and 
\[
\langle 0|\exp\left( \overline H(\overline t)+\hat H(\hat t)\right) v_i\hat v_j\exp\left(- \overline H(\overline t)-\hat H(\hat t)\right)|0\rangle =-\frac{\sqrt{-1}}2
a_{i,-N_i}a_{j,-N_j} s_{N_i}(\overline t+c_i)s_{N_j}(\hat t + c_j).
\]
Next, we calculate 
\[
\begin{aligned}
&\exp\left(  H( t)\right)v_i\exp\left(-  H( t)\right)=\\
&={\rm Res}\, \sum_{j=-N_i}^\infty\frac{ a_{ij}}{\sqrt 2}z^j \left(
 \exp\left(\sum_{k>0} t_k z^k\right)\psi^+(z)+z^{-1} \exp\left(\sum_{k>0} -t_k (-z)^k\right)\psi^-(-z)\right)dz\\
&=\frac{a_{i,-N_i}}{\sqrt 2}{\rm Res}\,  z^{-N_i}
 \left(
\exp\left(\sum_{k>0}( t_k+ c_{ik}) z^k\right)\psi^+(z)
+z^{-1}\exp\left(\sum_{k>0}((-1)^{k+1} t_k+ c_{ik}) z^k\right)\psi^-(-z)
\right)dz,
\end{aligned}
\]
and analogously, we find
\[
\begin{aligned}
&\exp\left(  H( t)\right)\hat v_i\exp\left(-  H( t)\right)=   
\frac{a_{i,-N_i}\sqrt{-1}}{\sqrt 2}{\rm Res}\,  z^{-N_i}\times\\
 &\left(
\exp\left(\sum_{k>0}( t_k+ c_{ik}) z^k\right)\psi^+(z)
-z^{-1}\exp\left(\sum_{k>0}((-1)^{k+1} t_k+ c_{ik}) z^k\right)\psi^-(-z)
\right)dz.
\end{aligned}
\]
From which we deduce that 
\[
\begin{aligned}
\langle& 0|\exp\left(  H( t)\right)v_iv_j\exp\left(-  H( t)\right)|0\rangle=\langle 0|\exp\left(  H( t)\right)\hat v_i\hat v_j\exp\left(-  H( t)\right)|0\rangle=\\
&=\frac{a_{i,-N_i}a_{j,-N_j}}{2}{\rm Res}\,  z^{-N_i}w^{-N_j}
\left(\sum_{k=0}^\infty \left(-\frac{w}{z}\right)^k
\exp\left(\sum_{k>0}( t_k+ c_{ik}) z^k+((-1)^{k+1} t_k+ c_{jk}) w^k\right)
\right.
\\
&\left. 
+\sum_{k=1}^\infty \left(-\frac{w}{z}\right)^k
\exp\left(\sum_{k>0}( t_k+ c_{jk}) w^k+((-1)^{k+1} t_k+ c_{ik}) z^k\right)
\right)\frac{dz}z\frac{dw}w\\
&=a_{i,-N_i}a_{j,-N_j}\chi^+_{N_i,N_j}(t+c_i,  t-\tilde c_j,t- \tilde c_i, t+c_j ).
\end{aligned}
\]
Here $\tilde c=(-c_1,c_2, -c_3,c_4,\ldots)$ and 
\begin{equation}
\label {chia}
\chi^\pm_{N,M}(s,t,u,v)= \frac12 \sum_{k=0}^{M} (-1)^{N}s_{N+k}(s)s_{M-k}(-t)\pm \frac12 \sum_{k=1}^{M} (-1)^M s_{N +k}(-u)s_{M-k}(v).
\end{equation}
Finally,
\[
\begin{aligned}
\langle 0|\exp&\left(  H( t)\right)v_i\hat v_j\exp\left(-  H( t)\right)|0\rangle=-\frac{\sqrt{-1}a_{i,-N_i}a_{j,-N_j}}{2}{\rm Res}\,  z^{-N_i}w^{-N_j}\times\\
&
\left(\sum_{k=0}^\infty \left(-\frac{w}{z}\right)^k
\exp\left(\sum_{k>0}( t_k+ c_{ik}) z^k+((-1)^{k+1} t_k+ c_{jk}) w^k\right)
\right.
\\
&\quad\left. 
-\sum_{k=1}^\infty \left(-\frac{w}{z}\right)^k
\exp\left(\sum_{k>0}( t_k+ c_{jk}) w^k+((-1)^{k+1} t_k+ c_{ik}) z^k\right)
\right)\frac{dz}z\frac{dw}w\\
=-&\sqrt{-1}a_{i,-N_i}a_{j,-N_j}\chi^-_{N_i,N_j}(t+c_i,  t-\tilde c_j,t- \tilde c_i, t+c_j ).
\end{aligned}
\]
\begin{remark}
If we multiply a KP or BKP tau-function by a non-zero scalar it is still a tau-function. Thus {\bf without loss of generality assume from now on  that all $a_{i, -N_i}=1$.} 
\end{remark}
In \cite{KvdLB2}, the following theorem was proved (cf. \cite{KRvdL}):
\begin{theorem} \label{Theo} All polynomial tau-functions of the BKP hierarchy are, up to a scalar
multiple, of the form
\begin{equation}
\label{TBKP}
\tau_B^{2n} (\overline t)=Pf\left(\overline \chi_{\lambda_i\lambda_j}(\overline t+c_i,\overline t+c_j)\right)_{1\le i,j\le 2n},
\end{equation}
where $\lambda=(\lambda_1,\lambda_2,\ldots ,\lambda_{2n}$ is an extended strict partition, i.e. $\lambda_1>\lambda_2>\cdots \lambda_{2n}\ge 0$, $\overline t=(t_1,0,t_3, 0,\ldots)$, $c_i=(c_{i1},c_{i2}, c_{i3}, \ldots)$ are constants and $\overline\chi_{N,M}$ is given by \eqref{chi}.
\end{theorem}
Note that formula \eqref{TBKP} is in fact $\hat\sigma (v_1v_2\cdots v_{2n}|0\rangle)$, for
\begin{equation}
\label{vvv}
\begin{aligned}
v_i&=\phi_{-\lambda_i} +\sum_{j>-\lambda_i}a_{ij}\phi_j\\
       &=\phi_{-\lambda_i} +\sum_{j>-\lambda_i}s_{j-\lambda_i}(c_i)\phi_j.
\end{aligned}
\end{equation}
The Pfaffian \eqref{TBKP} is obtained by applying  Wick's theorem.
Next, we calculate\break  $\hat\sigma (v_1v_2\cdots v_{2n}\hat v_1\hat v_2\cdots \hat v_{2n}|0\rangle)$. Using the above vacuum expectation values, we find that, with
\begin{equation}
\begin{aligned}
\overline A(\overline t)&=\left(\overline \chi_{\lambda_i\lambda_j}(\overline t+c_i,\overline t+c_j)\right)_{1\le i,j\le 2n},\\
\overline B(\overline t, \hat t)&=\left( -\frac{\sqrt{-1}}2 s_{\lambda_i}(\overline t+c_i)s_{\lambda_j}(\hat t + c_j)\right)_{1\le i,j\le 2n},
\end{aligned}
\end{equation}
the vacuum expectation value
\begin{equation}
\label{vvvvvv}
\begin{aligned}
\overline \sigma (v_1v_2\cdots v_{2n}\hat v_1\hat v_2\cdots \hat v_{2n}|0\rangle)&=
Pf\begin{pmatrix}\overline A(\overline t)&\overline B(\overline t, \hat t)\\
-\overline B(\overline t, \hat t)^T&\overline A(\hat  t)\end{pmatrix}\\
&=
Pf( \overline A(\overline t))
Pf(\overline A(\hat t))\\
&=\tau_B^{2n} (\overline t)\tau_B^{2n} (\hat t).
\end{aligned}
\end{equation}
In the second  equality of \eqref{vvvvvv}, we use  a formula, due to E. R. Caianiello \cite {C}, for a proof see \cite{O}. This formula expresses the Pfaffian of the  $4n\times 4n$ skew symmetric matrix 
in a sum of  determinants  times Pfaffians.  We first state the formula in more generality, than is needed here.
Denote by $[m]=\{ 1,2,\ldots, m\}$ for a nonnegative  integer $m$. Let  $I\subset [m]$ and $J\subset [k]$, denote by $W(I,J)$ the matrix that we construct out of  the 
$m\times n$ matrix $W$ by erasing the rows $i$ with  $i\in [m]\backslash I$ and the columns $j$ with $j\in [k]\backslash J$.
If $J=\{ j_1,j_2,\ldots, j_s\}$, we write $\sum(J)= j_1+j_2+\ldots+ j_s$. Then 
\begin{proposition}
Let $m$ and $k$ be nonnegative integers such that $m + k$ is even. Let $X$, respectively $Y$, be a skew-symmetric $m\times m$, respectively $k\times k$ matrix , and $W$ an arbitrary $m\times k$ matrix. Then
\begin{equation}
\label{Cai}
Pf
\begin{pmatrix}
X&W\\
-W^T&Y
\end{pmatrix}
=
\sum_{I,J}
\epsilon (I, J) Pf (X(I,I)) Pf( Y
(J,J)) \det (W([m] \backslash I; [k] \backslash J)),
\end{equation}
where the sum is taken over all pairs of even-element subsets $(I, J)$  such that $I \subset [m]$,
$J \subset [n]$ and where $m-|I|=k-|J|$.
Here
\[
\epsilon (I, J) = (-1)^{ \sum(I)+\sum(J)+{m\choose 2}+ 
{k\choose 2} +{{m-|I|}\choose 2}}.
\]
\end{proposition}
We use this proposition, to obtain the second  equality of \eqref{vvvvvv}. In our case $m=k=2n$ and  the matrix $W=\overline 
B(\overline t, \hat t)$ has rank 1, which means that all the terms on  the  right-hand side of \eqref{Cai} are zero, except when $I=J=[2n]$.

To obtain the Main Theorem of this paper, we calculate 
\[
 \sigma (v_1v_2\cdots v_{2n}\hat v_1\hat v_2\cdots \hat v_{2n}|0\rangle),
\]
From the above vacuum expectation values we deduce 
using Wick's theorem that the KP tau-function
\begin{equation}
\label{vvvkp}
\tau^{2n,2n}(t)= \sigma (v_1v_2\cdots v_{2n}\hat v_1\hat v_2\cdots \hat v_{2n}|0\rangle)=
Pf\begin{pmatrix} A^+(t)&-\sqrt{-1}A^-(t)\\\sqrt{-1}A^-(t)^T&A^+(t)\end{pmatrix},
\end{equation}
where (cf. \eqref{chia})
\begin{equation}
\label{AB}
A^\pm (t)=\left(\chi^\pm_{\lambda_i,\lambda_j}(t+c_i,  t-\tilde c_j, t-\tilde c_i, t+c_j )\right)_{1\le i,j\le 2n}.
\end{equation}
Thus we have shown:
\begin{theorem}
Let $\overline t=(\overline t_1,0,\overline t_3, 0,\ldots)$, then
the BKP tau-function $
\tau_B^{2n} (\overline t)$ of Theorem \ref{Theo}  is the square root of the KP tau-function $
\tau^{2n,2n}(t)$, given in \eqref{vvvkp}, i.e., 
\[
\tau_B^{2n} (\overline t)=\sqrt{\tau^{2n,2n}(\overline t)}.
\]
\end{theorem}
\begin{remark}If we put all the constants equal to zero, then the above KP tau-function $\tau^{2n,2n}(t)$  corresponds to the following element In $F$:
\[
\phi_{-\lambda_1}\phi_{-\lambda_2}\cdots\phi_{-\lambda_{2n}}\hat\phi_{-\lambda_1}\hat\phi_{-\lambda_2}\cdots\hat\phi_{-\lambda_{2n}}|0\rangle,
\]
which is up to a sign equal to
\[
\begin{cases}
\psi^+_{-\lambda_1+\frac12}\psi^+_{-\lambda_2+\frac12}\cdots\psi^+_{-\lambda_{2n}+\frac12}
\psi^-_{-\lambda_1-\frac12}\psi^-_{-\lambda_2-\frac12}\cdots\psi^-_{-\lambda_{2n}-\frac12}|0\rangle,
&
\mbox{ if }\lambda_{2n}\ne 0\ \mbox{and}\\
\psi^+_{-\lambda_1+\frac12}\psi^+_{-\lambda_2+\frac12}\cdots\psi^+_{-\lambda_{2n-1}+\frac12}
\psi^-_{-\lambda_1-\frac12}\psi^-_{-\lambda_2-\frac12}\cdots\psi^-_{-\lambda_{2n-1}-\frac12}|0\rangle,
&
\mbox{ if }\lambda_{2n}= 0.
\end{cases}
\]
This element corresponds to $s_{({\lambda_1-1},{\lambda_2-1},\cdots,{\lambda_{2n}-1}|{\lambda_1},{\lambda_2,}\cdots,{\lambda_{2n}})}(t)$, where we use  the Frobenius notation for a partition, see e.g. \cite{Mac}. This means that the tau-function $\tau^{2n,2n}(t)$ is the lowest element (i.e. it  generates,) the  KP Schubert cell  (cf. \cite{KvdLmodKP}) corresponding to the partition 
\[
\begin{cases}
({\lambda_1-1},{\lambda_2-1},\cdots,{\lambda_{2n}-1}|{\lambda_1},{\lambda_2,}\cdots,{\lambda_{2n}}),&
\mbox{ if }\lambda_{2n}\ne 0\ \mbox{and}\\
({\lambda_1-1},{\lambda_2-1},\cdots,{\lambda_{2n-1}-1}|{\lambda_1},{\lambda_2,}\cdots,{\lambda_{2n-1}}),
&
\mbox{ if }\lambda_{2n}= 0.
\end{cases}
\]
While  its "square root", the BKP tau-function  $
\tau_B^{2n} (\overline t)$, which is equal up to a multiplicative constant, to the Q-Schur function $ Q_{({\lambda_1},{\lambda_2,}\cdots,{\lambda_{2n}})}\left(\frac{t}2\right)$\begin{footnote}{One obtains the Q-Schur functions as given in Macdonald's book on symmetric functions \cite{Mac} by substituting for $t_i =2\sum_{j>0} \frac{x_j^i}i$.}\end{footnote}, 
generates the BKP Schubert cell corresponding to the strict partition
 $$({\lambda_1},{\lambda_2,}\cdots,{\lambda_{2n}}), \ \mbox{respectively }({\lambda_1},{\lambda_2,}\cdots,{\lambda_{2n-1}}), \ \mbox {if }\lambda_{2n}=0
.$$
\end{remark}
\section{The relation KdV versus BKP}
\label{S7}
A. Alexandrov showed in  a recent short publication \cite{Alex}, that all KdV tau-functions, i.e., KP tau-functions that are independen of the even times $t_{2i}$, are BKP tau-functions when one replaces all times $t_{2i+1}$ by   $\frac{t_{2i+1}}{2}$. In this section, we give a representation theoretical explanation for this.

It is clear from \cite{KvdLmodKP}, but was   proved already  in the 80's in \cite{KPeterson}, that all polynomial KdV tau functions can be obtained as the following vacuum expectation value, for a certain $k=0,1,2,\ldots$, where the $c_{2i-1}$ are arbitrary constants.
\begin{equation}
\label{KdV0}
\tau_k(t+c)
=\langle 0|e^{ \sum_{i=1}^\infty (t_{2i-1}+c_{2i-1})\alpha_{2i-1}}
\psi^+_{-k+\frac12}\psi^+_{-k+\frac52}\cdots \psi^+_{k-\frac12}|-k\rangle.
\end{equation}
Note that we can obtain all tau-functions \eqref{KdV1}, by calculating the above expression with all $c_{2i-1}=0$ and then substituting $t_{2i-1}+c_{2i-1}$ for $t_{2i-1}$. Thus from now on we will put all $c_{2i-1}=0$. In fact (cf.  \cite{KvdLmodKP} or \cite{KPeterson}),  it is not difficult to show that 
\[
\tau_k(t)=s_{(k,k-1,\ldots ,2,1)}(t_1,t_2,t_3,\dots)=s_{(k,k-1,\ldots ,2,1)}(t_1,0,t_3,0,\dots),
\]
the Schur function corresponding to the partition $\lambda={(k,k-1,\ldots ,2,1)}$, which is independent of the even times $t_{2i}$.
\begin{equation}
\label{KdV1}
\begin{aligned}
\tau_k(t)
&=\langle 0|e^{ \sum_{i=1}^\infty t_{2i-1}\alpha_{2i-1}}
\psi^+_{-k+\frac12}\psi^+_{-k+\frac52}\cdots \psi^+_{k-\frac12}|-k\rangle\\
&=\langle 0|e^{ \sum_{i=1}^\infty (t_{2i-1}+c_{2i-1})\alpha_{2i-1}} 
\psi^+_{-k+\frac12}\psi^+_{-k+\frac52}\cdots \psi^+_{k-\frac12}\psi^-_{-k+\frac12}\psi^-_{-k+\frac32}\cdots
\psi^-_{-\frac12}|0\rangle\\
&=\pm \langle 0|e^{ \sum_{i=1}^\infty t_{2i-1}\alpha_{2i-1}}
\psi^+_{-k+\frac12}\psi^-_{-k+\frac12}\psi^-_{-k+\frac52}\psi^-_{-k+\frac52}\cdots \psi^+_{-1-\frac{(-1)^k}2} \psi^-_{-1-\frac{(-1)^k}2}|0\rangle\\
&=\pm \langle 0|e^{ \sum_{i=1}^\infty t_{2i-1}\alpha_{2i-1}}
\psi^+_{-k+\frac12}\psi^-_{-k+\frac12}\psi^-_{-k+\frac52}\psi^-_{-k+\frac52}\cdots \psi^+_{-1-\frac{(-1)^k}2} \psi^-_{-1-\frac{(-1)^k}2}|0\rangle.\\
\end{aligned}
\end{equation}
The above calculation is up to a multiplicative sign. 
Now using \eqref{psiphi}, we can rewrite \eqref{KdV1}, again up to a sign, to:
\begin{equation}
\label{KdV2}
\begin{aligned}
\tau_k(t)&=\pm \langle 0|e^{ \sum_{i=1}^\infty t_{2i-1}\alpha_{2i-1}}
\frac{\phi_{-k}-\sqrt{-1}\hat  \phi_{-k}}{\sqrt 2}\frac{\phi_{-k+1}+\sqrt{-1}\hat  \phi_{-k+1}}{\sqrt 2}\times\\
&\qquad\qquad \frac{\phi_{-k+2}-\sqrt{-1}\hat  \phi_{-k+2}}{\sqrt 2}\dots
\frac{\phi_{ \frac{-1-(-1)^k}2}+\sqrt{-1}\hat  \phi_{ \frac{-1-(-1)^k}2}}{\sqrt 2}
|0\rangle .
\end{aligned}
\end{equation}
Instead of this expression, we focus on
\begin{equation}
\label{KdV3}
\begin{aligned}
\tau_k(s,t)&=\pm \langle 0|e^{ \sum_{i=1}^\infty s_{2i-1}\frac{\beta_{2i-1}}2+  t_{2i-1}\frac{\hat\beta_{2i-1}}2 }
\frac{\phi_{-k}-\sqrt{-1}\hat  \phi_{-k}}{\sqrt 2}\frac{\phi_{-k+1}+\sqrt{-1}\hat  \phi_{-k+1}}{\sqrt 2}\times\\
&\qquad\qquad \frac{\phi_{-k+2}-\sqrt{-1}\hat  \phi_{-k+2}}{\sqrt 2}\dots
\frac{\phi_{ \frac{-1-(-1)^k}2}+\sqrt{-1}\hat  \phi_{ \frac{-1-(-1)^k}2}}{\sqrt 2}
|0\rangle .
\end{aligned}
\end{equation}
Differentiate \eqref{KdV3} by $\frac{\partial}{\partial s_{2j+1}}-\frac{\partial}{\partial t_{2j+1}}$, we thus obtain
\begin{equation}
\label{KdV4}
\begin{aligned}
&\left(\frac{\partial}{\partial s_{2j+1}}-\frac{\partial}{\partial t_{2j+1}}\right)\tau_k(s,t)=\pm \langle 0|e^{ \sum_{i=1}^\infty s_{2i-1}\frac{\beta_{2i-1}}2+  t_{2i-1}\frac{\hat\beta_{2i-1}}2 }
\left(\frac{\beta_{2j-1}}2-\frac{\hat\beta_{2j-1}}{\sqrt 2} \right)\times\\
&\qquad\qquad\frac{\phi_{-k}-\sqrt{-1}\hat  \phi_{-k}}{\sqrt 2}
\frac{\phi_{-k+1}+\sqrt{-1}\hat  \phi_{-k+1}}{\sqrt 2}\dots
\frac{\phi_{ \frac{-1-(-1)^k}2}+\sqrt{-1}\hat  \phi_{ \frac{-1-(-1)^k}2}}{\sqrt 2}
|0\rangle .
\end{aligned}
\end{equation}
However, since
\[
\left[\frac{\beta_{2j-1}}2-\frac{\hat\beta_{2j-1}}2,
\frac{\phi_{-k+\ell}-(-1)^\ell\sqrt{-1}\hat  \phi_{-k+\ell}}{\sqrt 2} 
 \right]=\frac{\phi_{-k+\ell+2j-1}-(-1)^{\ell+2j-1}\sqrt{-1}\hat  \phi_{-k+\ell+2j-1}}{\sqrt 2},
\]
we conclude that
\[
\left(\frac{\partial}{\partial s_{2j+1}}-\frac{\partial}{\partial s_{2j+1}}\right)\tau_k(s,t)=0.
\]
Thus $\tau_k(s,t)$ is a function of $s+t$. From which we deduce that, up to a multiplicative sign,
\[
\pm\tau_k\left(\frac{t}2\right)= \tau_k\left(\frac{t}2,\frac{t}2\right)=\tau_k(\epsilon t,(1-\epsilon)t)=\tau_k(t,0).
\]
We now calculate explicitly $\tau_k(t,0)$.
Note first, that if $k$ is odd,
\[
\frac{\phi_0+\sqrt{-1}\hat\phi_0}{\sqrt 2}|0\rangle=\sqrt 2\phi_0|0\rangle.
\]
Thus
\begin{equation}
\label{KdV5}
\begin{aligned}
\tau_k(t,0)&=\pm \langle 0|e^{ \sum_{i=1}^\infty t_{2i-1}\frac{\beta_{2i-1}}2}
\frac{\phi_{-k}-\sqrt{-1}\hat  \phi_{-k}}{\sqrt 2}\frac{\phi_{-k+1}+\sqrt{-1}\hat  \phi_{-k+1}}{\sqrt 2}\times\\
&\qquad\qquad \frac{\phi_{-k+2}-\sqrt{-1}\hat  \phi_{-k+2}}{\sqrt 2}\dots
\frac{\phi_{ \frac{-1-(-1)^k}2}+\sqrt{-1}\hat  \phi_{ \frac{-1-(-1)^k}2}}{\sqrt 2}
|0\rangle\\
&=\pm\frac1{(\sqrt 2)^k} \langle 0|e^{ \sum_{i=1}^\infty t_{2i-1}\frac{\beta_{2i-1}}2}
\phi_{-k}\phi_{-k+1}\phi_{-k+2}\dots
\phi_{ \frac{-1-(-1)^k}2},
|0\rangle\\
\end{aligned}
\end{equation}
which clearly is a BKP tau-function. 

More explicitly, 
\begin{equation}
\label{KdV6}
\begin{aligned}
\tau_k(t,0)
&=\pm\frac1{(\sqrt 2)^k}  Pf\left(\overline \chi_{i,j}(t,t)\right)_{\frac{1-(-1)^k}2\le i,j\le k}\\
&=\pm\frac1{(\sqrt 2)^k} Q_{(k,k-1,\ldots ,2,1)} \left(\frac{t}2\right).
\end{aligned}
\end{equation}
Hence, it is (up to a multiplicative constant) the  Q-Schur function corresponding to the strict partition $(k,k-1,\ldots,2,1)$, and  it is, again up to a multiplicative  constant,  the square root of the Schur function $s_{(k^{k+1})}(t_1,0,t_3,0,\ldots)$.
We thus obtain (cf. \cite{Alex}):
\begin{proposition}
All polynomial  KdV tau-functions $\tau_k(t +c)=s_{(k,k-1,\ldots ,2,1)}(t_1+c_1,t_3+c_3,\ldots)$ become BKP tau-functions, when one replaces $t_i+c_i$ by $\frac{t_i+c_i}2$. Moreover, up to a multiplicative constant,  $\tau_k(\frac{t +c}2)$ is equal to 
$Q_{(k,k-1,\ldots ,2,1)} \left(\frac{t+c}2\right)$ and to the square root of
$s_{(k^{k+1})}(t_1+c_1,0,t_3+c_3,0,\ldots)$, here ${(k^{k+1})}$ is the partition $(k,k,\ldots,k)$, where $k$ appears $k+1$ times.
\end{proposition}

\end{document}